\pdfoutput=1
\RequirePackage{ifpdf}
\ifpdf % We are running pdfTeX in pdf mode
\documentclass[pdftex]{sigma}
\else
\documentclass{sigma}
\fi

\numberwithin{equation}{section}

\begin{document}

\allowdisplaybreaks

\newcommand{\arXivNumber}{1704.00043}

\renewcommand{\thefootnote}{}

\renewcommand{\PaperNumber}{053}

\FirstPageHeading

\ShortArticleName{Symmetries of the Hirota Dif\/ference Equation}

\ArticleName{Symmetries of the Hirota Dif\/ference Equation\footnote{This paper is a~contribution to the Special Issue on Symmetries and Integrability of Dif\/ference Equations. The full collection is available at \href{http://www.emis.de/journals/SIGMA/SIDE12.html}{http://www.emis.de/journals/SIGMA/SIDE12.html}}}

\Author{Andrei K.~POGREBKOV~$^{\dag\ddag}$}

\AuthorNameForHeading{A.K.~Pogrebkov}

\Address{$^\dag$~Steklov Mathematical Institute of Russian Academy of Science, Moscow, Russia}
\EmailD{\href{mailto:pogreb@mi.ras.ru}{pogreb@mi.ras.ru}}
\URLaddressD{\url{http://www.mathnet.ru/eng/person13595}}

\Address{$^\dag$~National Research University Higher School of Economics, Moscow, Russia}
\URLaddressD{\url{https://www.hse.ru/en/org/persons/66265278}}

\ArticleDates{Received March 31, 2017, in f\/inal form July 02, 2017; Published online July 07, 2017}

\Abstract{Continuous symmetries of the Hirota dif\/ference equation, commuting with shifts of independent variables, are
derived by means of the dressing procedure. Action of these symmetries on the dependent variables of the equation is
presented. Commutativity of these symmetries enables interpretation of their parameters as ``times'' of the nonlinear
integrable partial dif\/ferential-dif\/ference and dif\/ferential equations. Examples of equations resulting in such procedure
and their Lax pairs are given. Besides these, ordinary, symmetries the additional ones are introduced and their action on
the Scattering data is presented.}


\Classification{35Q51; 37K10; 37K15; 37K40; 39A14}

\renewcommand{\thefootnote}{\arabic{footnote}}
\setcounter{footnote}{0}

\section{Introduction}

Hirota dif\/ference equation was introduced in the bilinear form (HBDE) as equation on the $\tau$-function
in~\cite{Hirota1977,Hirota1981},
\begin{gather}
\tau^{(1)}(n)\tau^{(2,3)}(n)+\tau^{(2)}(n)\tau^{(3,1)}(n)+\tau^{(3)}(n)\tau^{(1,2)}(n)=0,\label{hbde}
\end{gather}
where $\tau(n)=\tau(n_1,n_2,n_3)$ is a function of 3 numbers (independent variables) $n_1,n_2,n_3\in\mathbb{Z}$. Here and below upper indexes 1, 2, 3 in parenthesis denote unity shifts $\tau^{(i)}(n)=\tau(n)|_{n_i\to n_i+1}$, matrix $\tau^{(i,j)}$ is antisymmetric and $\tau^{(i,j)}(n)=\tau(n)_{n_i\to n_i+1, n_j\to n_j+1}$ for $1\leq{i}<{j}\leq3$. HBDE has a lot of literature since it is known to generate by means of special limiting procedures many discrete and continuous integrable equations, such as Kadomtsev--Petviashvili equation (KP), modif\/ied Kadomtsev--Petviashvili equation, two-dimensional Toda lattice, sine-Gordon equation, Benjamin--Ono equation, etc. Because of this HDE is often considered to be a fundamental integrable system. This equation also appears as the model-independent functional relation for eigenvalues of quantum transfer matrices. Detailed survey of the results referred to this equation is given in~\cite{Zabrodin1997,Zabrodin2008}, see also citations therein. Octahedral structure of HDE is studied in \cite{Saito}. Its elliptic solutions were considered in~\cite{Krichever/Wiegmann/Zabrodin}. In~\cite{Bogdanov/Konopelchenko} Hirota dif\/ference equation is called the generalized KP hierarchy. Some Hirota-like equations are derived in \cite{Fioravanti/Nepomechie}. Here we use the form of the Hirota dif\/ference equation that is convenient and natural for the Inverse scattering transform, but that is dif\/ferent from HBDE in (\ref{hbde}). We introduce a~``f\/ield variable'', i.e., function $v(n)=v(n_1,n_2,n_3)$ def\/ined by means of equalities
\begin{gather}
v^{(1)}(n)-v^{(3)}(n)=\frac{\tau^{(3,1)}(n)\tau(n)}{\tau^{(1)}(n)\tau^{(3)}(n)}, \qquad
v^{(2)}(n)-v^{(1)}(n)=\frac{\tau^{(1,2)}(n)\tau(n)}{\tau^{(2)}(n)\tau^{(1)}(n)},\label{vtau1}
\end{gather}
where in analogy to the above we use notations
\begin{alignat}{3}
& v^{(1)}(n)=v(n_1+1,n_2,n_3),\qquad && v^{(2)}(n)=v(n_1,n_2+1,n_3), \quad \text{etc.},& \label{5}\\
&v^{(11)}(n)=v(n_1+2,n_2,n_3),\qquad && v^{(12)}(n)=v(n_1+1,n_2+1,n_3), \quad \text{etc.}, & \label{6}
\end{alignat}
but in contrast to the $\tau^{(i,j)}$ matrix $v^{(i,j)}$ is symmetric. Summing up equations in (\ref{vtau1}) we get by~(\ref{hbde})
 \begin{gather}
v^{(3)}(n)-v^{(2)}(n)=\frac{\tau^{(2,3)}(n)\tau(n)}{\tau^{(2)}(n)\tau^{(3)}(n)}.\label{vtau2}
\end{gather}
Notice that all these three equations follow consequently by cyclic permutation of the indexes $\{1,2,3\}$. Now, by (\ref{vtau1}) we have that $\big(v^{(2)}-v^{(1)}\big)^{(3)}\big(v^{(3)}-v^{(1)}\big)=\big(v^{(3)}-v^{(1)}\big)^{(2)}\big(v^{(2)}-v^{(1)}\big)$, or
\begin{gather}
v^{(12)}\big(v^{(2)}-v^{(1)}\big)+v^{(23)}\big(v^{(3)}-v^{(2)}\big)+v^{(31)}\big(v^{(1)}-v^{(3)}\big)=0,\label{n19}
\end{gather}
that is the form of the Hirota dif\/ference equation (HDE) used below. This equation is known (see, e.g.,~\cite{Zabrodin1997}) to be the compatibility condition for the Lax pair given by any pair of the equations
\begin{gather}
\varphi^{(i)}=\varphi^{(j)}+\big(v^{(i)}-v^{(j)}\big)\varphi,\qquad i,j=1,2,3.\label{59'}
\end{gather}
It is worth to mention that because of rational dependence on $\tau$ of the r.h.s.'s of~(\ref{vtau1}) and~(\ref{vtau2}) equations~(\ref{hbde}) and~(\ref{n19}) describe dif\/ferent classes of solutions, see \cite{Pogrebkov2011}. Say, the f\/irst one does not support solutions such that some of the dif\/ferences in the l.h.s.'s annihilates at some~$n$. On the other side, equation~(\ref{n19}) is ill posed as it is: any function~$v(n)$ such that
\begin{gather}
v^{(i)}=v^{(j)} \qquad \text{for}\quad  i\neq{j} \label{ill}
\end{gather}
obeys this equation. Below we derive additional conditions that resolve this problem.

Here we use the HDE to demonstrate a generic approach to construction of symmetries of integrable equations. In~\cite{Pogrebkov2011} it was shown that the HDE can be derived by dressing of a commutator identity on an associative algebra. More exactly, let us have some associative algebra with unity over the complex space $\mathbb{C}$. Let us f\/ix some complex mutually dif\/ferent parameters~$a_1$, $a_2$ and $a_3$. It is easy to see, that for any pair $A$, $B$ of elements of this algebra, such that in this algebra there exist~$(A-a_1)^{-1}$, $ (A-a_2)^{-1}$ and $(A-a_3)^{-1}$, we have the following identity:
\begin{gather}
a_{12} \big\{(A-a_1)(A-a_2)B(A-a_1)^{-1}(A-a_2)^{-1}+(A-a_3)B(A-a_3)^{-1}\big\}\nonumber\\
\qquad {}+\text{cycle}(1,2,3)=0,\label{1}
\end{gather}
where
\begin{gather}
 a_{ij}=a_{i}-a_{j},\qquad a_{i}\neq a_{j}\qquad\text{for}\quad i\neq{j}.\label{8}
\end{gather}
Due to a mutual commutativity of elements $A-a_i$, we can introduce dependence of $B$ on discrete ``times'' $n_i\in\mathbb{Z}$, $i=1,2,3$, by means of equation
\begin{gather}
 B(n_1,n_2,n_3)=\left(\prod_{i=1}^{3}(A-a_i)^{n_i}\right)B\left(\prod_{i=1}^{3}(A-a_i)^{n_i}\right)^{-1}.\label{4}
\end{gather}
Denoting for shortness $B(n)=B(n_1,n_2,n_3)$ we use notation of the kind (\ref{5}) and (\ref{6}). In these terms (\ref{1}) proves that $B(n)$ obeys the following dif\/ference equation
\begin{gather}
a_{12}\big\{B^{(12)}+B^{(3)}\big\}+\text{cycle}(1,2,3)=0,\label{7}
\end{gather}
that is a linearized version of the HDE (\ref{n19}), as we explain after (\ref{n18}) below. The special dressing procedure that enables such ``nonlinearization'' of equation~(\ref{7}) was introduced in~\cite{Pogrebkov2011}, where we also described derivation of integrable dif\/ferential and dif\/ference-dif\/ferential equations from HDE in limiting cases. Direct and inverse problems for the HDE were considered in~\cite{Pogrebkov2014}. Our construction here is based on the Abelian version of the HDE for simplicity. See~\cite{Pogrebkov2016} for analogous treatment of the non-Abelian HDE.

The article is organized as follows. In Section~\ref{dress} we shortly present some steps of the dressing procedure necessary for our approach, see \cite{Pogrebkov2011} for details. The symmetries and their dressing are introduced and studied in Section~\ref{symm}. In Section~\ref{eqs} we list some integrable equations and corresponding Lax pairs that appear by consideration of the parameters of symmetries as independent variables. Additional symmetries and concluding remarks are presented in Section~\ref{add}.

\section{Dressing procedure}\label{dress}

In order to introduce dressing procedure we f\/ix representation of associative algebra. We realize it as set of operators~$F$,~$G$, etc., given by their symbols $\widetilde{F}(n_1,z)$, $\widetilde{G}(n_1,z)$, etc., being functions of the discrete variable $n_1\in\mathbb{Z}$ and $z\in\mathbb{C}$. We assume that these symbols have Fourier transform
\begin{gather}
\widetilde{F}(n_1,z)=\oint_{|\zeta|=1}\frac{d\zeta}{2\pi i\zeta}\zeta^{n_1}f(\zeta,z),\label{f1}
\end{gather}
where $f(\zeta,z)$ is a function (distribution) of its variables, $\zeta,z\in\mathbb{C}$, $|\zeta|=1$. Notation of the kind $f(\zeta,z)$ does not mean that any analyticity property of symbols with respect to $z$ are assumed. So there exists nontrivial operation of~$\overline\partial$-dif\/ferentiation on this set of operators, i.e., to any operator $F$ we associate operator $\overline\partial{F}$ with symbol
\begin{gather}
\widetilde{(\overline{\partial}F)}(n_1,z)=\frac{\partial \widetilde{F}(n_1,z)}{\partial\overline{z}},\label{38}
\end{gather}
where derivative is considered in the sense of distributions. As well on this set of operators we def\/ine composition
\begin{gather}
\widetilde{(FG)}(n_1,z)=\oint_{|\zeta|=1}\frac{d\zeta}{2\pi i\zeta}\widetilde{F}(n_1,z\zeta) \sum_{m_1\in\mathbb{Z}}\zeta^{n_1-m_1}\widetilde{G}(m_1,z),\label{fg}
\end{gather}
if it exists. It is easy to check associativity of this composition law.

As the simplest examples let us consider unity and shift operators. The symbol of the unity (with respect to composition~({\ref{fg}})) operator is~1, and symbol of the operator~$T$ that shifts variable $n_1$ equals
\begin{gather*}
\widetilde{T}(n_1,z)=z,%\label{331}
\end{gather*}
so that by (\ref{fg}) for any $F$:
\begin{gather}
\widetilde{(TF)}(n,z)=z\widetilde{F^{(1)}}(n,z),\qquad \widetilde{\big(FT^{-1}\big)}(n,z)=\frac{1}{z}\widetilde{F}(n,z),\nonumber\\
\widetilde{\big(TFT^{-1}\big)}(n,z)=\widetilde{F^{(1)}}(n,z),\label{41}
\end{gather}
where notation (\ref{5}) was used. We see that symbol of operator $T$ is analytic, so that by~(\ref{38})
\begin{gather}
 \overline\partial{T}=0,\label{40}
\end{gather}
that is very essential for the following construction. It is easy to see that any operator $F$ with symbol $\widetilde{F}(n,z)=f(|z|)$, i.e., independent of~$n_1$ and~$\arg{z}$, commute with an arbitrary operator in the sense of composition~(\ref{33}).

Below we also use a multiplication operator $N_{1}$ with symbol $\widetilde{N_1}(n_1,z)=n_1$. Due to~(\ref{fg}) it is conjugate to operator $T$ in a~sense that
\begin{gather}
 [T,N_1]=T,\label{fg1}
\end{gather}
and for a generic $F$
\begin{gather}
 \widetilde{[N_1,F]}(n_1,z)=i\frac{\partial\widetilde{F}(n_1,z)}{\partial\arg{z}}.\label{fg2}
\end{gather}

We consider now element $B$ in (\ref{4}) to be an operator of this class, i.e., given by the symbol
\begin{gather}
\widetilde{B}(n_1,z)=\oint_{|\zeta|=1}\frac{d\zeta}{2\pi i\zeta}\zeta^{n_1}b(\zeta,z),\label{f2}
\end{gather}
with some function $b(\zeta,z)$, cf.~(\ref{f1}). If compare (\ref{4}) and~(\ref{41}) it is natural to set
\begin{gather}
 A=T_1+a_1,\qquad\text{i.e.},\qquad \widetilde{A}(n,z)=z_1+a_1,\label{44}
\end{gather}
so that shift of $n_1$ in (\ref{4}) (see notation (\ref{5})) gives
\begin{gather}
B^{(1)}=TBT^{-1},\label{n8}
\end{gather}
that is valid for any operator due to (\ref{41}). Specif\/ic for $B$ is that dependence on discrete variables~$n_2$ and~$n_3$ is given by~(\ref{4}) and~(\ref{44}):
\begin{gather}
B^{(2)}(n_2,n_3)\equiv B(n_2+1,n_3) =(T+a_{12})B(n_2,n_3)(T+a_{12})^{-1},\label{n9}\\
B^{(3)}(n_2,n_3)\equiv B(n_2,n_3+1)=(T+a_{13})B(n_2,n_3)(T+a_{13})^{-1}.\label{n10}
\end{gather}
that we denote below as $B(n)$ and the symbol as $\widetilde{B}(n_1,n_2,n_3,z)$ ($\widetilde{B}(n,z)$ for shortness). For operators which symbols depend on all three discrete variables the composition law~(\ref{fg}) takes the form
\begin{gather}
\widetilde{(FG)}(n,z)=\oint_{|\zeta|=1}\frac{d\zeta}{2\pi i\zeta}\widetilde{F}(n,z\zeta) \sum_{m_1\in\mathbb{Z}}\zeta^{n_1-m_1}\widetilde{G}(m_1,n_2,n_3,z).\label{33}
\end{gather}

Thanks to this equation, (\ref{f2}), (\ref{n9}) and (\ref{n10}) the symbol of the operator $B(n)$ equals
\begin{gather}
\widetilde{B}(n,z)=\oint_{|\zeta|=1}\frac{d\zeta}{2\pi i\zeta}\zeta^{n_1}\left(\frac{z\zeta+a_{12}}{z+a_{12}}\right)^{n_2}
\left(\frac{z\zeta+a_{13}}{z+a_{13}}\right)^{n_3}b(\zeta,z).\label{b1}
\end{gather}
It is reasonable to exclude its exponential growth with respect to $n_2$ and $n_3$. So we impose conditions $|z\zeta+a_{12}|=|z+a_{12}|$, $|z\zeta+a_{13}|=|z+a_{13}|$, that are equivalent to either $\zeta=1$, or  $\overline{z}/{z} =\zeta\overline{a}_{12}/a_{12} =\zeta\overline{a}_{13}/a_{13}$. The f\/irst condition leads to a trivial constant operator in (\ref{b1}), so we consider the second one only. Because of it: $\overline{a}_{12}/a_{12}=\overline{a}_{13}/a_{13}$, and thus (shifting phase of~$z$, if necessary) we can choose all $a_j$ to be real. This means that function $b(\zeta,z)$ has support on the surface $\zeta=\overline{z}/z$. In the simplest case $b(\zeta,z)= \delta_{c}(\zeta{z}/\bar{z})\widetilde{R}(z)$, where $\delta_{c}$ is the $\delta$-function on the unit circle and $\widetilde{R}(z)$ is an arbitrary function of $z\in\mathbb{C}$. Then representation~(\ref{b1}) for the symbol of~$B$ becomes
\begin{gather}
\widetilde{B}(n,z)=\left(\frac{\overline{z}}{z}\right)^{n_1}\left(\frac{\overline{z}+a_{12}}{z+a_{12}}\right)^{n_2}
 \left(\frac{\overline{z}+a_{13}}{z+a_{13}}\right)^{n_3}\widetilde{R}(z).\label{n11}
\end{gather}
Taking property of the $n$-dependent factor here into account, it is reasonable to input condition that $\widetilde{R}(\overline{z})= \overline{\widetilde{R}(z)}$. Then also $\widetilde{B}(n,\overline{z})=\overline{\widetilde{B}(n,z)}$. In generic situation $b(\zeta,z)$ in (\ref{b1}) can be proportional to the f\/inite sum of derivatives of $\delta_{c}(\zeta)$, that we do not consider here in order to avoid asymptotic growth of $\widetilde{B}(n,z)$ by~$n$.

The dressing procedure is based on construction of the dressing operator $K(n)$ by means of the $\overline\partial$-problem
\begin{gather}
 \overline\partial{K}(n)=K(n)B(n),\qquad \lim_{z\to\infty}\widetilde{K}(n,z)=1,\label{49}
\end{gather}
where $\widetilde{K}(n,z)$ is the symbol of operator $K(n)$. Evolutions (\ref{n8})--(\ref{n10}) generate evolutions of the dressing operator: $\overline\partial{K}^{(j)}(n)=K^{(j)}(n)B^{(j)}(n)$, where we use notation~(\ref{5}) for~$K(n)$. In order to describe these shifts we need to specify asymptotic condition in~(\ref{49}) assuming decomposition
\begin{gather}
\widetilde{K}(n,z)=\sum_{j=0}^{M}k_{j}(n)z^{-j}+o\big(z^{-M}\big),\qquad z\to\infty,\qquad k_{0}(n)\equiv1,\label{n24}
\end{gather}
where $M$ is some f\/inite positive number and functions $k_j(n)$ are independent of $z$, i.e., are operators that coincide with their symbols. By (\ref{40}) and (\ref{n8})--(\ref{n10}) we have that $\overline\partial{K}^{(j)}(T+a_{1j})={K}^{(j)}(T+a_{1j})B$, $j=1,2,3$ and $a_{11}=0$ by~(\ref{8}). Thus ${K}^{(j)}(T+a_{1j})$ for any $j$ obeys the same equation as in (\ref{49}) but its symbol grows linearly at $z$-inf\/inity. Assuming unique solvability of the problem~(\ref{49}) we see that there exist operators $P_{j}$ with symbols being entire functions of $z$, such that ${K}^{(j)}(T+a_{1j})=P_{j}K$. Asymptotic decomposition~(\ref{n24}) shows that symbols $\widetilde{P_{j}}(n,z)$ are polynomials of the f\/irst order with respect to $z$. Their coef\/f\/icients follow from the latter equality and~(\ref{n24}). Thus $P_{1}=T$, so that $K^{(1)}=TKT^{-1}$ as it must be for any operator under consideration. But results for shifts with respect to the second and third discrete variables are less trivial
\begin{gather}
K^{(2)}(T+a_{12})=K^{(1)}T+\bigl[u^{(2)}-u^{(1)}+a_{12}\bigr]K,\label{55}\\
K^{(3)}(T+a_{13})=K^{(1)}T+\bigl[u^{(3)}-u^{(1)}+a_{13}\bigr]K,\label{56}
\end{gather}
where we denoted for simplicity
\begin{gather}
u=k_1\label{54}
\end{gather}
(see (\ref{n24})), so we consider $u$ as operator with symbol $\widetilde{u}(n,z)=u(n)$. Equations~(\ref{55}) and~(\ref{56}) are dressed versions of equations
\begin{gather}
B^{(2)}(T+a_{12})=B^{(1)}T+a_{12}B,\qquad B^{(3)}(T+a_{13})=B^{(1)}T+a_{13}B,\label{50}
\end{gather}
that follow from (\ref{n8})--(\ref{n10}). By construction, evolutions of $K$ with respect to the discrete ``times'' are compatible and compatibility of (\ref{55}) and (\ref{56}) gives
\begin{gather}
 u^{(12)}(u^{(2)}-u^{(1)}+a_{12})+a_{12}u^{(3)}+\text{cycle}(1,2,3)=0,\label{57}
\end{gather}
that is the Hirota dif\/ference equation on the function $u=u(n_1,n_2,n_3)$. It is clear that~(\ref{7}) is a linearization of this equation.

Substituting decomposition (\ref{n24}) in (\ref{55}) and (\ref{56}) one def\/ines coef\/f\/icients $k_{j}$ of this decomposition. In this way we get
\begin{gather*}
\begin{split}
& k_{j+1}^{(2)}+a_{12}k_{j}^{(2)}=k_{j+1}^{(1)}+\big(u_{}^{(2)}-u_{}^{(1)}+a_{12}^{}\big)k_{j}^{},\\ %\label{n:28}\\
& k_{j+1}^{(3)}+a_{13}k_{j}^{(3)}=k_{j+1}^{(1)}+\big(u_{}^{(3)}-u_{}^{(1)}+a_{13}^{}\big)k_{j}^{},\qquad j\geq0.%\label{n:29}
\end{split}
\end{gather*}
In particular,
\begin{gather}
k_{2}^{(2)}-k_{2}^{(1)}=\big(u_{}^{(2)}-u_{}^{(1)}\big)u-a_{12}^{}\big(u_{}^{(2)}-u\big),\label{n:30}\\
k_{2}^{(3)}-k_{2}^{(1)}=\big(u_{}^{(3)}-u_{}^{(1)}\big)u-a_{13}^{}\big(u_{}^{(3)}-u\big),\label{n:31}\\
k_{3}^{(2)}-k_{3}^{(1)}=-a_{12}^{}k_{2}^{(2)}+\big(u_{}^{(2)}-u_{}^{(1)}+a_{12}^{}\big)k_{2},\label{n:32}\\
k_{3}^{(3)}-k_{3}^{(1)}=-a_{13}^{}k_{2}^{(3)}+\big(u_{}^{(3)}-u_{}^{(1)}+a_{13}^{}\big)k_{2}.\label{n:33}
\end{gather}
These relations are nonlocal and need assumptions on asymptotic behavior of coef\/f\/icients $k_{j}(n)$ to be uniquely solvable.

In order to simplify the above relations we introduce the Jost solution as
\begin{gather}
\varphi(n,z)=z^{n_1}(z+a_{12})^{n_2}(z+a_{13})^{n_3}\widetilde{K}(n,z),\label{58:1}
\end{gather}
Then equations (\ref{55}) and (\ref{56}) of the Lax pair take the form
\begin{gather}
\varphi^{(2)}=\varphi^{(1)}+\big(u^{(2)}-u^{(1)}+a_{12}\big)\varphi,\label{59} \\
 \varphi^{(3)}=\varphi^{(1)}+\big(u^{(3)}-u^{(1)}+a_{13}\big)\varphi.\label{60}
\end{gather}
It is also worth to mention that the dif\/ference of these equations gives
\begin{gather}
\varphi^{(3)}=\varphi^{(2)}+\big(u^{(3)}-u^{(2)}+a_{23}\big)\varphi,\label{61}
\end{gather}
that is also symmetric with respect to the previous equations. Thus any two of these three equations can be taken as the Lax pair. Let us perform change of dependent variable
\begin{gather}
v(n)=u(n)-a_1n_1-a_2n_2-a_3n_3,\label{n18}
\end{gather}
where by construction $u(n)\to0$ when $n\to\infty$. Thus by (\ref{5}) and (\ref{6}) $v^{(i)}-v^{(j)}=u^{(i)}-u^{(j)}+a_{ji}$, so by (\ref{n18}) we get HDE (\ref{n19}) and its Lax pair~(\ref{59'}) from~(\ref{57}) and (\ref{59})--(\ref{61}). Moreover, the condition (\ref{8}) and asymptotic behavior of $v(n)$ in~(\ref{n18}) resolve the ill posedness of~(\ref{n19}), see discussion of~(\ref{ill}) in Introduction.

\section{Continuous symmetries of the Hirota dif\/ference equations}\label{symm}

In the case where nonlinear equation under consideration is integrable, i.e., has nontrivial Lax pair, construction of symmetries of the equation is equivalent to construction of the Lax pair, here (\ref{59}), (\ref{60}). Because of existence of the inverse problem such direct procedure, being rather complicated by itself, can be substituted by the following two steps (see~\cite{Zakharov/Manakov}, where analogous approach was used for construction of integrable equations, and~\cite{Pogrebkov2000}). Taking that the Lax pair (\ref{59}), (\ref{60}) (or, strictly speaking, (\ref{55}), (\ref{56})) appeared as dressing of the ``bare'' pair (\ref{50}), we start with construction of symmetries of (\ref{50}). Then symmetries of the HDE itself are given by the dressing procedure, i.e., by the inverse problem (\ref{49}). Because of (\ref{n8})--(\ref{n10}) the simplest set of symmetries of the pair (\ref{50}) is given by operators that commute with operator $T$, i.e., functions of operator $T$ itself. Explicitly, we introduce dependence of $B$ on a new (continuous) variable $t$ by means of relation
\begin{gather}
 B_t(n,t)=[W,B(n,t)],\label{n200}
\end{gather}
where symbol of the operator $W$ equals $\widetilde{W}(n,z)=w(z)$, being a meromorphic function of $z$. Because of the composition law (\ref{33}) any such operator commutes with operator~$T$, thus thanks to~(\ref{n8})--(\ref{n10}):
\begin{gather}
(B_{t})^{(j)}=\big(B^{(j)}\big)_{t},\label{n2010}
\end{gather}
where parenthesis denote order of operations. Moreover, let us have an operator $W'$ of the same kind and introduce dependence on $t'$ in analogy to (\ref{n200}). Then $[W,W']=0$ and these symmetries commute: $\partial_{t'}\partial_{t}B(n,t,t')=\partial_{t'}\partial_{t}B(n,t,t')$. In terms of symbols we have thanks to~(\ref{b1}) that
\begin{gather*}
\widetilde{B}(n,t,z)=\oint_{|\zeta|=1}\frac{d\zeta}{2\pi i\zeta}\zeta^{m_1}\left(\frac{z\zeta+a_{12}}{z+a_{12}}\right)^{m_2}
\left(\frac{z\zeta+a_{13}}{z+a_{13}}\right)^{m_3}e_{}^{t(w(z\zeta)-w(z))}b(\zeta,z),%\label{n201}
\end{gather*}
that preserves form of $\widetilde{B}$, redef\/ining $b(\zeta,z)$ only. Our next step is to def\/ine dressing operators by means of~(\ref{49}), so that $\overline\partial{K}_{t}(n,t)=K_{t}(n,t)B(n,t)+K(n,t)B_{t}(n,t)$, or thanks to~(\ref{n200}):
\begin{gather}
 \overline\partial{K}_{t}(n,t)+K(n,t)B(n,t)W=\big(K_{t_n}(n,t)+K(n,t)W\big)B(n,t).\label{n202}
\end{gather}
Below we consider special cases of such symmetries that enables explicit integration of this equation due to (\ref{40}).

\subsection{Symmetries of the KP type}
This set of symmetries is generated by the simplest choices of operator $W$ in~(\ref{n200}): $w(z)=z,z^{2},\ldots$. In other words, we introduce dependence of operator $B$ on the times $t_1,t_2,\ldots$ by means of the equations $B_{t_m}=[T^{m},B]$, $m=1,2,\ldots$, and def\/ine dependence of the dressing operator on these times by means of (\ref{n202}): $\overline\partial{K}_{t_m}+KBT^{m}=(K_{t_m}+KT^{m})B$. Thanks to~(\ref{40}) this equation is equivalent to
\begin{gather}
 \overline\partial\big({K}_{t_m}+KT^{m}\big)=\big({K}_{t_m}+KT^{m}\big)B.\label{n21}
\end{gather}
We see that the sum ${K}_{t_m}+KT^{m}$ obeys the same $\overline\partial$-equation as in~(\ref{49}), but with dif\/ferent (polynomial with respect to $z$) asymptotic of symbol at $z$-inf\/inity. Thus for any $m$ there exists operator $P_{m}$ such that $\overline{\partial}P_{m}=0$ and
\begin{gather}
 {K}_{t_m}+KT^{m}=P_{m}K.\label{n22}
\end{gather}
Assuming that asymptotic in~(\ref{49}) is dif\/ferentiable, we see that symbol $\widetilde{P}_{m}(n,t,z)$ is polynomial of the $m$-th power with respect to~$z$. Coef\/f\/icients~$p_{m,j}$ of this polynomial,
\begin{gather}
P_{m}=\sum_{j=0}^{m}p_{m,j}T^{j},\label{n:22}
\end{gather}
where symbols $\widetilde{p}_{m,j}(n,t)$ depend on~$n$ and~$t$, but not on~$z$, are def\/ined (in analogy to the standard Zakharov--Shabat dressing procedure, \cite{Zakharov/Shabat}) by equality
\begin{gather}
\big(\widetilde{K}(n,t,z)z^{m}\big)_{+}=\big((\widetilde{P_{m}K})(n,t,z)\big)_{+},\label{n23}
\end{gather}
where $+$ denotes polynomial by $z$ part of symbols of operators.

Under substitution of (\ref{n24}), where now coef\/f\/icients $k_{j}$ depend on $t_n$, in (\ref{n23}) we get recursion relations for coef\/f\/icients in (\ref{n:22}):
\begin{gather*}
 k_{m-m'}=\sum_{i=m'}^{m}p_{m,i}k_{i-m'}^{(1\times{i})},%\label{n:23}
\end{gather*}
where $k_{m}^{(1\times{i})}(n_1,n_2,n_3,t)=k_{m}^{}(n_1+i,n_2,n_3,t)$ in correspondence to (\ref{5}). In particular, $p_{m,m}(t,n,z)\equiv1$ and for the three lowest symmetries we have explicitly
\begin{gather}
P_{1}=T+k_{1}^{}-k_{1}^{(1)}\equiv T+u-u^{(1)},\label{n25}\\
P_{2}=T^{2}+\big(k_{1}^{}-k_{1}^{(11)}\big)T+k_{2}^{}-k_{2}^{(11)}-\big(k_{1}^{}-k_{1}^{(11)}\big)k_{1}^{(1)},\label{n26}\\
P_{3}=T^{3}+\big(k_{1}^{}-k_{1}^{(111)}\big)T^{2}+\big(k_{2}-k_{2}^{(111)}-\big(k_{1}^{}-k_{1}^{(111)}\big)k_{1}^{(11)}\big)T\nonumber\\
\hphantom{P_{3}=}{}
+k_{3}^{}-k_{3}^{(111)}-\big(k_{2}^{}-k_{2}^{(111)}-\big(k_{1}^{}-k_{1}^{(111)}\big) k_{1}^{(11)}\big)k_{1}^{(1)}+\big(k_{1}^{(111)}-k_{1}^{}\big)k_{2}^{(11)},\label{n27}
\end{gather}
where for the upper indexes in parenthesis we use notation in (\ref{5}), (\ref{6}).

Notice that action of the f\/irst symmetry on the dressing operator is given in terms of dependent variable~$u$ of the HDE~(\ref{57}):
\begin{gather}
K_{t_1}=\big(K^{(1)}-K\big)T+\big(u-u^{(1)}\big)K,\label{n28}
\end{gather}
but action of this symmetry on the $u=k_{1}$ itself involves $k_{2}$. Indeed, by (\ref{n24}) $1/z$ term of~(\ref{n28}) gives
\begin{gather}
u_{t_1}=k_{2}^{(1)}-k_{2}^{}-\big(u^{(1)}-u\big)u,\label{n29}\\
k_{2,t_1}=k_{3}^{(1)}-k_{3}^{}-\big(u^{(1)}-u\big)k_{2},\label{n291}
\end{gather}
and so on. Coef\/f\/icient $k_{2}$ in (\ref{n24}) is given by (\ref{n:30}), (\ref{n:31}), so that action of this symmetry on~$u$ is nonlocal.

Coef\/f\/icients of polynomials (\ref{n25})--(\ref{n27}) take a much simpler form if we write them in terms of $u$ and its derivatives with respect to~$t_m$. Thus, for the second symmetry, i.e., (\ref{n22}) for $m=2$, we get by~(\ref{n26}) and derivative of~(\ref{n28}) by~$t_1$: $K_{t_2}-K_{t_1t_1}=\big(P_2^{}-P_{1}^{2}-P_{1,t_1}^{}\big)K+2K_{t_1}T$. But thanks to (\ref{n25}) and (\ref{n26}), (\ref{n29}) $P_2^{}-P_{1}^{2}-P_{1,t_1}^{}=-2u_{t_1}$, so that
\begin{gather*}
 K_{t_2}-K_{t_1t_1}-2K_{t_1}T=-2u_{t_1}K.%\label{n300}
\end{gather*}

Next, thanks to (\ref{n27}) we derive that
\begin{gather}
 K_{t_3}-K_{t_1t_1t_1}=\big(P_{3}-P_{1}^{3}-2P_{1,t_1}P_{1}-P_{1}P_{1,t_1}-P_{1,t_1t_1}\big)K +3(P_{1}K)_{t_1}T.\label{n314}
\end{gather}
In order to simplify the r.h.s.\ we use (\ref{n29}) and (\ref{n291}) to derive
\begin{gather*}
k_{2}^{}-k_{2}^{(111)}-\big(k_{1}^{}-k_{1}^{(111)}\big)k_{1}^{(11)}\\
\qquad{} =-\big(u+u_{}^{(1)}+u_{}^{(11)}\big)_{t_1} +\big(u^{(11)}-u^{(1)}\big)^{ 2 } +\big(u^{(1)}-u\big)\big(u^{(11)}-u\big),\\ %\label {n312}\\
k_{3}^{(111)}-k_{3}^{}=\big(k_{2}^{}+k_{2}^{(1)}+k_{2}^{(11)}\big)_{t_1} \\
\hphantom{k_{3}^{(111)}-k_{3}^{}=}{} +\big(u^{(1)}-u\big)k_{2}^{}+\big(u^{(11)}-u^{(1)}\big)k_{2}^{(1)}+\big(u^{(111)}-u^{(11)}\big)k_{2}^{(11)},%\label{n313}
\end{gather*}
that gives f\/inally
\begin{gather}
P_{3}-P_{1}^{3}-2P_{1,t_1}P_{1}-P_{1}P_{1,t_1}-P_{1,t_1t_1}=-3u_{t_1}T-\frac{3}{2}u_{t_1t_1}- \frac{3}{2}u_{t_2}.\label{n315}
\end{gather}

It is clear that in the same way symmetries corresponding to the highest times $t_{m}$ can be studied. Specif\/ic property of all these symmetries is the analyticity (polynomiality) of operator $W$ that allowed us to rewrite~(\ref{n202}) in the form~(\ref{n21}) and to control asymptotic behavior of~$P_{m}$ by means of~(\ref{n22}).

\subsection{Singular symmetries}
Here we consider symmetries def\/ined by operators $W$ with symbols $w(z)$ being meromorphic functions of~$z$. Arbitrary symmetries of this kind need consideration based on the object more generic than the Jost solution, the so called Cauchy--Jost function, see~\cite{Boiti/Pempinelli/Pogrebkov,Grinevich/Orlov}. So here we consider only the simplest examples of such symmetries: those where $W$ is chosen to be one of operators $(T+a_{1j})^{-1}$, see notation (\ref{8}) (in particular, $a_{11}=0$). We introduce dependence of operator $B$ on the set $t_{-}=\{t_{-1}, t_{-2}$, $t_{-3}\}$ of three continuous parameters by means of relations (cf.~(\ref{n200}))
\begin{gather}
 B_{t_{-j}}(n,t_{-})=\big[(T+a_{1j})^{-1},B(n,t_{-})\big],\qquad j=1,2,3.\label{n35}
\end{gather}
Then (\ref{n202}) gives $\overline\partial{K}_{t_{-j}}+KB(T+a_{1j})^{-1}=K_{t_{-j}}B+K(T+a_{1j})^{-1}B$. Multiplying this equation by $(T+a_{1j})$ from
the right we get
\begin{gather*}
 \overline\partial\bigl(K_{t_{-j}}(T+a_{1j})+K\bigr)=\bigl(K_{t_{-j}}(T+a_{1j})+K\bigr)B^{(-j)},%\label{n36}
\end{gather*}
where $B^{(-j)}=(T+a_{1j})^{-1}B(T+a_{1j})$, cf.\ (\ref{n8}), i.e.,
\[
 \widetilde{B}^{(-j)}(n_1,n_2,n_3,t_{-},z)=\widetilde{B}(n_1,n_2,n_3,t_{-},z)\bigr|_{n_j\to n_j-1},\qquad j=1,2,3.
\]
Taking now (\ref{49}) into account, we see that there exist operators $V_{j}$, such that
\begin{gather}
 K_{t_{-j}}(n,t_{-})(T+a_{1j})+K(n,t_{-})=V_{j}(n,t_{-})K^{(-j)}(n,t_{-}).\label{n37}
\end{gather}
In terms of the symbols this equation sounds as
\[
(z+a_{1j})\widetilde{K}_{t_{-j}}(n,t_{-},z)+\widetilde{K}(n,t_{-},z)=\big(\widetilde{V_{j}K}\big)^{(-j)}(n,t_{-},z),
\]
and thanks to the asymptotic condition in (\ref{49}) symbols of operators $V_{j}$ do not depend on $z$, so that $(\widetilde{V_{j}K})^{(-j)}(n,t_{-},z)=\widetilde{V_{j}}(n,t_{-})\widetilde{K}^{(-j)}(n,t_{-},z)$. In spite of the singular behavior of the symbols of operators $(T+a_{1j})^{-1}$, we are not interested here in transformations that change spectrum of the Lax operators, so the dressing operator $K$ and its derivative~$K_{t_{-j}}$ have no pole singularities with respect to~$z$. Thus setting~$z=a_{j1}$ in the previous equality, we derive for the symbol of operator~$V_{j}$:
\begin{gather}
\widetilde{V_{j}}(n,t_{-})=\frac{\widetilde{K}(n,t_{-},a_{j1})}{\widetilde{K}^{(-j)}_{}(n,t_{-},a_{j1})}.\label{n38}
\end{gather}
This relation is highly nonlocal in terms of coef\/f\/icients~$k_{j}$ of the expansion~(\ref{n24}). On the other side action of this symmetry on these coef\/f\/icients is easily given by means of~(\ref{n37}). Say, thanks to~(\ref{54})
\begin{gather}
u_{t_{-j}}=V_{j}-1.\label{n39}
\end{gather}
In the same way other ``singular'' symmetries can be introduced and their action on the dressing operator and solution of the HDE can be derived.

\subsection{Formulation in terms of the Jost solutions}
In order to simplify the above results, we redef\/ine the Jost solution (cf.~(\ref{58:1})) as
\begin{gather}
\varphi(n,t, t_{-},z)=z^{n_1}(z+a_{12})^{n_2}(z+a_{13})^{n_3} \nonumber\\
 \hphantom{\varphi(n,t, t_{-},z)=}{} \times e^{t_1z+t_2z^{2}+t_3z^{3}+t_{-1}z^{-1}+t_{-2}(z+a_ { 12 } )^ { -1 }+t_{-3}(z+a_{13})^{-1}} \widetilde{K}(n,t,t_{-},z).\label{n32}
\end{gather}
In this way we do not change equations (\ref{59})--(\ref{61}) and the above results read as follows. Instead of (\ref{n28}) we have
\begin{gather}
\varphi_{t_1}=\varphi^{(1)}+\big(u-u^{(1)}\big)\varphi.\label{n33}
\end{gather}
Next, by (\ref{n22}) for $n=2$ and (\ref{n25}) we reduce (\ref{n26}) to
\begin{gather}
 \varphi_{t_2}=\varphi^{(11)}+\big(u-u^{(11)}\big)\varphi^{(1)}+\big(k_{2}^{}-k_{2}^{(11)}-\big(u-u^{(11)}\big)u^{(1)}\big)\varphi,\label{n331}
\end{gather}
where $k_2$ must be def\/ined by~(\ref{n:30}). So in terms of the HDE variables $n_1$, $n_2$ and $n_3$ this symmetry is nonlocal. On the other side, taking $t_1$-dependence into account we derive by (\ref{n29}) that $k_2^{}-k_2^{(11)}=-u^{}_{t_1}-u^{(1)}_{t_1}-\big(u^{(1)}-u\big) u-\big(u^{(11)}-u^{(1)}\big)u^{(1)}$, so that (\ref{n331}) takes the form
\begin{gather}
\varphi_{t_2}=\varphi^{(11)}+\big(u-u^{(11)}\big)\varphi^{(1)}+\big(\big(u^{(1)}-u\big)^{2}-u^{}_{t_1}-u^{(1)}_{t_1}\big)\varphi.\label{n332}
\end{gather}
Finally, due to (\ref{n33}) we reduce this equality to more simple and familiar form:
\begin{gather}
\varphi_{t_2}=\varphi_{t_1t_1}-2u_{t_1}\varphi.\label{n34}
\end{gather}
Analogously, action of the symmetry (\ref{n27}) in terms of the shifts of the discrete variables is given by expression that involves~$k_3$. But in terms of the $t_1$-derivatives equation (\ref{n314}) reduces to
\begin{gather}
\varphi_{t_3}=\varphi_{t_1t_1t_1}-3u_{t_1}\varphi_{t_1}-\frac{3}{2}u_{t_1t_1}\varphi -\frac{3}{2}u_{t_2}\varphi\label{n40}
\end{gather}
due to (\ref{n315}). It is necessary to mention that direct check that the above relations give symmetries of the HDE involves relations (\ref{54}), (\ref{n:30})--(\ref{n:33}) and (\ref{n29}), (\ref{n291}), (and some their consequences) and is rather cumbersome. In our approach commutativity of these symmetries with the HDE evolution follows from simple relations of the linear case.

Notice also that under substitution (\ref{n32}) symmetry (\ref{n37}) reduces to
\begin{gather}
 \varphi_{t_{-j}}=V_{j}\varphi^{(-j)},\label{n41}
\end{gather}
where $V_{j}=1+u_{t_{-j}}$ by (\ref{n39}). Compatibility of this symmetry with the Lax pair~(\ref{59}) and~(\ref{60}) follows by~(\ref{n38}), (\ref{n39}), if we take into account that due to~(\ref{55}) and~(\ref{56})
\begin{gather*}
a_{1j}\widetilde{K^{(j)}}(n,t,t_{-},0)=\big[u^{(j)}-u^{(1)}+a_{1j}\big]\widetilde{K}(n,t,t_{-},0),\qquad j=1,2,3.%\label{n42}
\end{gather*}

\section{Evolution equations generated by the standard symmetries}\label{eqs}
All symmetries introduced above, see (\ref{n33})--(\ref{n41}), are compatible with HDE evolution, i.e., with any pair of equations (\ref{59})--(\ref{61}). Moreover, these symmetries mutually commute by construction. Thus it is natural to assume that the discrete variables~$n_1$, $n_2$, $n_3$, and continuous ones $t_1$, $t_2$, $t_3$, $t_{-1}$, $t_{-2}$, $t_{-3}$, etc., can be considered equally in a sense that any three of them can be chosen as independent variables of a~(dif\/ferential, or dif\/ferential-dif\/ference) nonlinear integrable equation, with a Lax pair given by a proper choice from the list of all equations \mbox{(\ref{59})--(\ref{61})} and (\ref{n33})--(\ref{n41}). Then other variables serve as (discrete, or continuous) para\-me\-ters of symmetries for this nonlinear equation. This assumption is obvious if the variable~$n_1$ is taken as one of independent variables, while needs special proof of closeness in a generic case. Here we consider some equations that result from the derived above symmetries of the HDE.

\looseness=-1 Say, we see that (\ref{n34}), (\ref{n40}) is nothing but the Lax pair of the KPII equation (see \cite{Dryuma, Kadomtsev/Petviashvili1970, Zakharov/Shabat}) for $2u_{t_1}$, where $t_1$, $ t_2$ and $t_3$ are the standard KP times. Then relations \mbox{(\ref{59})--(\ref{61})} and~(\ref{n33}),~(\ref{n41}) give the discrete and continuous symmetries of the KPII equation. The discrete symmetries are nothing but the B{\"a}cklund transformations of the~KPII. Some intermediate equations can be obtained in the following way. Let us choose variables~$n_1$,~$n_2$ and~$t_1$, and correspondingly let the Lax pair be given by~(\ref{59}) and~(\ref{n33}). Compatibility of these two equations gives{\samepage
\begin{gather}
 \frac{\partial}{\partial t_1}\log\big(u^{(2)}-u^{(1)}+a_{12}\big)+u^{(12)}-u^{(2)}-u^{(1)}+u=0,\label{n44}
\end{gather}
or $\partial_{t_1}\log\big(v^{(2)}-v^{(1)}\big)+v^{(12)}-v^{(2)}-v^{(1)}+v=0$ in notation (\ref{n18}).}

Another possibility is to choose variables $n_1$, $n_2$ and $t_2$ and the Lax pair given by~(\ref{59}) and~(\ref{n331}). Then corresponding nonlinear integrable equation reads as
\begin{gather}
 \frac{\partial}{\partial t_2}\log(u^{(2)}-u^{(1)}+a_{12}) +\big[k_{2}^{(11)}-k_{2}^{} -(u^{(11)}-u^{})u^{(1)}\big]^{(2)} \nonumber\\
\qquad {} -\big[k_{2}^{(11)}-k_{2}^{}-(u^{(11)}-u^{})u^{(1)}\big]=0,\label{n45}
\end{gather}
where $k_2$ is def\/ined by~(\ref{n:30}). Thus this equation is nonlocal and it can be considered as the higher analog of the equation~(\ref{n44}). Both these equations give continuous symmetries of the HDE, while evolution with respect to $n_3$ due to HDE evolution gives B{\"a}cklund transformation for both these nonlinear equations.

Let us consider now choice of variables $n_1$, $t_1$ and $t_2$ with Lax pair given by~(\ref{n33}) and~(\ref{n332}). It is easy to check that condition of compatibility of these two equations reads as
\begin{gather}
 \big(u^{(1)}-u\big)_{t_2}-\big(u^{(1)}+u\big)_{t_1t_1}+\big(u^{(1)}-u\big)^{2}=0.\label{n46}
\end{gather}
These equation can be considered as continuous symmetry of the (\ref{n44}) with respect to the variable $t_2$, or as continuous symmetry of (\ref{n45}) with respect to the variable~$t_1$. On the other side both~(\ref{n44}) and~(\ref{n45}) gives B{\"a}cklund transformations of~(\ref{n46}), i.e., shift of the variable~$n_2$.

The same is valid for the singular symmetries introduced in~(\ref{n35}). Taking~(\ref{n41}) as one of equations of the Lax pair we have in the r.h.s.\ shift of variable $n_j$. So we choose from \mbox{(\ref{59})--(\ref{61})} the equation that gives a shift with respect to some other variable $n_i$, $i\neq{j}$, in terms of the $n_j$-shift:
\begin{gather}
\varphi^{(i)}=\varphi^{(j)}+\big(u^{(i)}-u^{(j)}+a_{ji}\big)\varphi,\qquad i,j=1,2,3,\qquad i\neq{j}.\label{n47}
\end{gather}
Compatibility of this equation with (\ref{n35}) gives two conditions, one of which is identity due to~(\ref{n39}) and the another reduces to
\begin{gather}
 \frac{(1+u_{t_{-j}})^{(i)}}{1+u_{t_{-j}}}=\frac{u^{(i)}-u^{(j)}+a_{ji}}{(u^{(i)}-u^{(j)}+a_{ji})^{(-j)}},\label{n48}
\end{gather}
that is action of the $t_{-j}$-symmetry on the dependent variable of HDE. On the other side, taking~(\ref{n38}) and~(\ref{n39}) into account we rewrite this equality as
\begin{gather*}
\frac{u^{(i)}-u^{(j)}+a_{ji}}{(u^{(i)}-u^{(j)}+a_{ji})^{(-j)}}=
\frac{\widetilde{K}^{(i)}(a_{j1})\widetilde{K}^{(-j)}(a_{j1})}{\widetilde{K}^{(i,-j)}(a_{j1})\widetilde{K}(a_{j1})},%\label{n49}
\end{gather*}
that proves independence of the ratio $\big(u^{(i)}-u^{(j)}+a_{ji}\big)\widetilde{K}(a_{j1})/\widetilde{K}^{(i)}(a_{j1})$ on $n_j$. Assuming that $u(n,t_{-})$ decays and $\widetilde{K}(n,t_{-},z)\to1$ when $n\to\infty$, we get
\begin{gather}
u^{(i)}(n,t_{-})-u^{(j)}(n,t_{-})+a_{ji}=a_{ji}\frac{\widetilde{K}^{(i)}(n,t_{-},a_{j1})}{\widetilde{K}(n,t_{-},a_{j1})},\label{n50}
\end{gather}
i.e., relation of $u$ with the nonlocal objects in the r.h.s. The antisymmetry of the l.h.s.\ with respect to $i$ and $j$
shows that
\begin{gather*}
\frac{\widetilde{K}^{(i)}(n,t_{-},a_{j1})}{\widetilde{K}(n,t_{-},a_{j1})}=
\frac{\widetilde{K}^{(j)}(n,t_{-},a_{i1})}{\widetilde{K}(n,t_{-},a_{i1})},\qquad
\text{for any} \quad i,j=1,2,3.%\label{n51}
\end{gather*}
Further on, dif\/ferentiating (\ref{n50}) with respect to $t_{-j}$ and denoting $\widetilde{K}(n,t_{-},a_{j1})=e^{\psi(n,t_{-})}$ we get by means of (\ref{n39}) integrable equation
\begin{gather*}
 a_{ji}\big(\psi^{(i)}-\psi\big)_{t_{-j}}=e^{\psi-\psi^{(i,-j)}}-e^{\psi^{(j)}-\psi^{(i)}},%\label{n52}
\end{gather*}
with respect to the independent variables $n_i$, $n_j$ and $t_{-j}$. Its Lax pair follows from (\ref{n41}) and~(\ref{n47}) under above substitutions
\begin{gather*}
 \varphi^{(i)}=\varphi^{(j)}+a_{ji}e^{\psi^{(i)}-\psi}\varphi,\qquad %\label{n53}\\
 \varphi_{t_{-j}}=e^{\psi-\psi^{(-j)}}\varphi^{(-j)}.%\label{n54}
\end{gather*}

In analogy we can consider an equation with one discrete and two continuous variables from the set $t_{-}$. We
choose $n_1$, $t_{-1}$ and $t_{-2}$ and start from (\ref{n41}) for $j=1,2$. Let, for shortness,
$v(n,t_-)=u(n,t_-)-a_{1}n_1-a_{2}n_2+t_{-1}+t_{-2}$, cf.\ (\ref{n18}). Thanks to (\ref{n39}) this enables us to rewrite
 (\ref{n41}) as
\begin{gather}
 \varphi^{(1)}_{t_{-1}}=v^{(1)}_{t_{-1}}\varphi,\qquad \varphi^{(2)}_{t_{-2}}=v^{(2)}_{t_{-2}}\varphi.\label{n55}
\end{gather}
The f\/irst equation here goes as one of equations of the corresponding Lax pair, and we have to exclude shift with respect
to the variable $n_2$ in order to get the second equation. For this aim we dif\/ferentiate (\ref{59}) with respect to
$t_{-2}$ and substituting $\varphi^{(2)}_{t_{-2}}$ by (\ref{n55}) we get
\begin{gather*}
\varphi^{(1)}_{t_{-2}}=w\varphi_{t_{-2}}+v^{(1)}_{t_{-2}}\varphi,%\label{n56}
\end{gather*}
i.e., the second equation of the Lax pair. Here we denoted $w=v^{(1)}-v^{(2)}$. Compatibility condition now sounds as
\begin{gather*}
\partial_{t_{-2}}\log{v}_{t_{-1}}=
\frac{v_{t_{-2}}}{w^{(-1)}}-\biggl(\frac{v_{t_{-2}}}{w^{(-1)}}\biggr)^{(1)},\qquad %\label{n57}\\
\partial_{t_{-1}}\log{w}=\biggl(\frac{v_{t_{-1}}}{w^{(-1)}}\biggr)^{(1)}-\frac{v_{t_{-1}}}{w^{(-1)}}, %\label{n58}
\end{gather*}
where the f\/irst equation can be considered as the evolution one on the function~$v(n,t_{-})$, while the second stands as def\/inition of the auxiliary function~$w(n,t_{-})$.

Finally, let us notice that symmetries of the set $t_{-}$ can be combined with the KP type symmetries due to their mutual commutativity. As an example we choose variables~$t_{1}$,~$n_j$ and~$t_{-j}$ for some $j=1,2,3$. Taking that (\ref{n41}) includes shift with respect to the~$n_j$ (i.e., $\varphi^{(-j)}$) into account, we use (\ref{59}), (\ref{60}) to exclude~$\varphi^{(1)}$ from~(\ref{n33}):
\begin{gather*}
 \varphi_{t_1}=\varphi^{(j)}+\big(u-u^{(j)}+a_{1j}\big)\varphi.%\label{n65}
\end{gather*}
Then due to (\ref{n39}) condition of compatibility of this equation with (\ref{n41}) gives
\begin{gather}
 \partial_{t_1}\log(1+u_{t_{-j}})=2u-u^{(j)}-u^{(-j)}.\label{n66}
\end{gather}
This nonlinear integrable equation can be considered also as continuous symmetry of (\ref{n48}). Dif\/ferentiating (\ref{n66})
by $t_{-j}$ we use (\ref{n39}) to write the result in the form
$\partial_{t_1}\partial_{t_{-j}}\log{V}_{j}=2V_{j}^{}-V_{j}^{(j)}-V_{j}^{(-j)}$. Under substitution
$V_{j}=e^{\psi^{(j)}-\psi}$ we conclude that dif\/ference
$\partial_{t_1}\partial_{t_{-j}}\log\psi-e^{\psi^{(j)}-\psi}-e^{\psi-\psi^{(-j)}}$ is independent of $n_j$. Thus, under the
same assumption as above on the asymptotic decay of $u$ (and then of $\psi$) with respect to $n_j$, we get equation of the
2-dimensional Toda chain,
\begin{gather*}
\partial_{t_1}\partial_{t_{-j}}\psi=e^{\psi^{(j)}-\psi}-e^{\psi-\psi^{(-j)}}.%\label{n67}
\end{gather*}
It is clear that this list of integrable equations that appear as obeying (\ref{n2010}) symmetries of the HDE can be continued.

\section{Additional symmetries}\label{add}

In Section~\ref{symm} we mentioned that symmetries of the HDE appear as dressing of symmetries that preserve the bare pair~(\ref{50}). In this sense symmetries considered above are too restrictive. Indeed, condition that operator $W$ in~(\ref{n200}) has symbol depending on the variable $z$ only leads to~(\ref{n2010}), i.e., to commutativity of these symmetries with shifts of the independent variables of the HDE. This condition is enough, but not necessary in order to preserve (\ref{n50}). Notice, that if some operator~$X$ obeys the same relations~(\ref{50}) as operator~$B$, then thanks to~(\ref{n8}) also $BX$ and $[X,B]$ obey these relations. As an example of operator $X$ we take~(\ref{n11}) with $\widetilde{R}(z)\equiv1$. Thanks to~(\ref{40}) we have that~$\overline\partial{X}$ also obeys~(\ref{50}). Finally, we introduce operator $N=(\overline\partial{X})X^{-1}T$, symbol of which equals
\begin{gather*}
\widetilde{N}(n,z)=n_1+n_2+n_3-\frac{a_{12}n_2}{z+a_{12}}-\frac{a_{13}n_3}{z+a_{13}}.%\label{a1}
 \end{gather*}
By (\ref{33}) it is easy to check, that this operator can be written as
 \begin{gather*}
 N(n_2,n_3)=(T+a_{12})^{n_2}(T+a_{13})^{n_3}N_{1}(T+a_{12})^{-n_2}(T+a_{13})^{-n_3},%\label{a:1}
 \end{gather*}
where operator $N_1$ was considered in (\ref{fg1}), (\ref{fg2}). Thus, in analogy to (\ref{n200}) we introduce dependence of operator $B$ on a new variable $s\in\mathbb{R}$ by means of equation
\begin{gather}
 B(n,s)=e^{isN}B(n)e^{-isN},\qquad B(n,0)=B(n).\label{a2}
\end{gather}
Operator $N$ does not commute with shift operators in (\ref{n8})--(\ref{n10}): it is easy to see that
\begin{gather*}
 [T,N]=T,\qquad j=1,2,3,%\label{a3}
\end{gather*}
and thanks to (\ref{b1}) we get in terms of operator $B(0)=B(n_2,n_3)|_{n_2=n_3=0}$
\begin{gather}
 B(n,s)=(T+a_{12})^{n_2}(T+a_{13})^{n_3}e^{isN_1}B(0)e^{-isN_1}(T+a_{12})^{-n_2}(T+a_{13})^{-n_3},\label{a:4}
\end{gather}
or explicitly for the symbol of operator $B(n,s)$:
\begin{gather}
\widetilde{B}(n,s,z)=\oint_{|\zeta|=1}\frac{d\zeta}{2\pi i\zeta}\zeta^{n_1} \left(\frac{z\zeta+a_{12}}{z+a_{12}}\right)^{n_2}\left(\frac{z\zeta+a_{13}}{z+a_{13}}\right)^{n_3}b\big(\zeta,ze^{-is}\big).\label{a4}
\end{gather}
The above mentioned noncommutativity results in the following observation. Def\/ining in analogy to (\ref{a2})
\begin{gather*}
 B^{(j)}(n,s)=e^{isN}B^{(j)}(n)e^{-isN},\label{a5}
 \end{gather*}
 we have
\begin{gather}
 B^{(j)}(n,s)=\big(Te^{-is}+a_{1j}\big)B(n,s)\big(Te^{-is}+a_{1j}\big)^{-1},\label{a6}
 \end{gather}
 while by (\ref{a:4}), or (\ref{a4})
\begin{gather}
B(n,s)^{(j)}=(T+a_{1j})B(n,s)(T+a_{1j})^{-1},\label{a7}
\end{gather}
that coincides with (\ref{a6}) for $j=1$ only ($a_{11}=0$ by (\ref{8})). On the other side, multiplying (\ref{a7}) by $(T+a_{1j})$ we get that operator $B(n,s)$ obeys (\ref{n8})--(\ref{n10}) for all $j=1,2,3$.

Thus we have an example of additional symmetry of the kind considered for the nonlinear Schr\"{o}dinger equation in~\cite{OS} and for the KPI in~\cite{Pogrebkov2000}. Dependence of the corresponding dressing operator $K(n,s)$ on $s$ must be def\/ined again by the $\overline\partial$-problem~(\ref{49}) with operator $B(n,s)$, but this consideration we postpone for the forthcoming article. Dif\/ferential equations generated by this symmetry also deserve special consideration. It is interesting to mention that opera\-tor~$N$ can be taken as operator~$A$ in~(\ref{4}). So it also generate some discrete integrable equation, since identity~(\ref{1}) is valid for any operators~$A$ and~$B$. It is worth to mention that relation with HDE of some of equations obtained above is well known. Some of them appear as continuous limits of HDE, some as symmetries, see, e.g.,~\cite{Pogrebkov2011,Zabrodin1997}. Here we used an example of the HDE to develop a generic scheme of construction of the symmetries, both ordinary and additional, that can be applied to any integrable nonlinear equation.

\subsection*{Acknowledgements}

This work has been funded by the Russian Academic Excellence Project `5-100'.

\pdfbookmark[1]{References}{ref}
\LastPageEnding


\begin{thebibliography}{99}
\footnotesize\itemsep=-0.5pt

\bibitem{Bogdanov/Konopelchenko}
Bogdanov L.V., Konopelchenko B.G., Generalized {KP} hierarchy: {M}\"obius
 symmetry, symmetry constraints and {C}alogero--{M}oser system,
 \href{https://doi.org/10.1016/S0167-2789(01)00161-0}{\textit{Phys.~D}} \textbf{152/153} (2001), 85--96, \href{https://arxiv.org/abs/solv-int/9912005}{solv-int/9912005}.

\bibitem{Boiti/Pempinelli/Pogrebkov}
Boiti M., Pempinelli F., Pogrebkov A.K., Cauchy--{J}ost function and hierarchy
 of integrable equations, \href{https://doi.org/10.1007/s11232-015-0367-y}{\textit{Theo\-ret. and Math. Phys.}} \textbf{185}
 (2015), 1599--1613, \href{https://arxiv.org/abs/1508.02229}{arXiv:1508.02229}.

\bibitem{Dryuma}
Dryuma V.S., Analytic solution of the two-dimensional {K}orteweg--de {V}ries
 {(KdV)} equation, \textit{JETP Lett.} \textbf{19} (1974), 387--388.

\bibitem{Fioravanti/Nepomechie}
Fioravanti D., Nepomechie R.I., An inhomogeneous {L}ax representation for the
 {H}irota equation, \href{https://doi.org/10.1088/1751-8121/aa5303}{\textit{J.~Phys.~A: Math. Theor.}} \textbf{50} (2017),
 054001, 14~pages, \href{https://arxiv.org/abs/1609.06761}{arXiv:1609.06761}.

\bibitem{Grinevich/Orlov}
Grinevich P.G., Orlov A.Yu., Virasoro action on {R}iemann surfaces,
 {G}rassmannians, {$\det \overline\partial_J$} and {S}egal--{W}ilson
 {$\tau$}-function, in Problems of Modern Quantum Field Theory ({A}lushta,
 1989), \href{https://doi.org/10.1007/978-3-642-84000-5_7}{\textit{Res. Rep. Phys.}}, Springer, Berlin, 1989, 86--106.

\bibitem{Hirota1977}
Hirota R., Nonlinear partial dif\/ference equations. {II}.~{D}iscrete-time {T}oda
 equation, \href{https://doi.org/10.1143/JPSJ.43.2074}{\textit{J.~Phys. Soc. Japan}} \textbf{43} (1977), 2074--2078.

\bibitem{Hirota1981}
Hirota R., Discrete analogue of a generalized {T}oda equation, \href{https://doi.org/10.1143/JPSJ.50.3785}{\textit{J.~Phys.
 Soc. Japan}} \textbf{50} (1981), 3785--3791.

\bibitem{Kadomtsev/Petviashvili1970}
Kadomtsev B.B., Petviashvili V.I., On the stability of solitary waves in weakly
 dispersive media, \textit{Sov. Phys. Dokl.} \textbf{192} (1970), 539--541.

\bibitem{Krichever/Wiegmann/Zabrodin}
Krichever I., Wiegmann P., Zabrodin A., Elliptic solutions to dif\/ference
 non-linear equations and related many-body problems, \href{https://doi.org/10.1007/s002200050333}{\textit{Comm. Math.
 Phys.}} \textbf{193} (1998), 373--396, \href{https://arxiv.org/abs/hep-th/9704090}{hep-th/9704090}.

\bibitem{OS}
Orlov A.Yu., Shul'man E.I., Additional symmetries of the nonlinear Schr\"odinger
 equation, \href{https://doi.org/10.1007/BF01017968}{\textit{Theoret. and Math. Phys.}} \textbf{64} (1985), 862--866.

\bibitem{Pogrebkov2000}
Pogrebkov A.K., On time evolutions associated with the nonstationary
 {S}chr\"odinger equation, in L.{D}.~{F}addeev's {S}eminar on {M}athematical
 {P}hysics, \href{https://doi.org/10.1090/trans2/201/13}{\textit{Amer. Math. Soc. Transl. Ser.~2}}, Vol.~201, Amer. Math.
 Soc., Providence, RI, 2000, 239--255, \href{https://arxiv.org/abs/math-ph/9902014}{math-ph/9902014}.

\bibitem{Pogrebkov2011}
Pogrebkov A.K., Hirota dif\/ference equation and a commutator identity on an
 associative algebra, \href{https://doi.org/10.1090/S1061-0022-2011-01153-7}{\textit{St.~Petersburg Math.~J.}} \textbf{22} (2011),
 473--483.

\bibitem{Pogrebkov2014}
Pogrebkov A.K., Hirota dif\/ference equation: inverse scattering transform,
 {D}arboux transformation, and solitons, \href{https://doi.org/10.1007/s11232-014-0237-z}{\textit{Theoret. and Math. Phys.}}
 \textbf{181} (2014), 1585--1598, \href{https://arxiv.org/abs/1407.0677}{arXiv:1407.0677}.

\bibitem{Pogrebkov2016}
Pogrebkov A.K., Commutator identities on associative algebras, the
 non-{A}belian {H}irota dif\/ference equation and its reductions,
 \href{https://doi.org/10.1134/S0040577916060039}{\textit{Theoret. and Math. Phys.}} \textbf{187} (2016), 823--834.

\bibitem{Saito}
Saito S., Octahedral structure of the {H}irota--{M}iwa equation,
 \textit{J.~Nonlinear Math. Phys.} \textbf{19} (2012), 539--550.

\bibitem{Zabrodin1997}
Zabrodin A.V., Hirota dif\/ference equations, \href{https://doi.org/10.1007/BF02634165}{\textit{Theoret. and Math. Phys.}}
 \textbf{113} (1997), 1347--1392, \mbox{\href{https://arxiv.org/abs/solv-int/9704001}{solv-int/9704001}}.

\bibitem{Zabrodin2008}
Zabrodin A.V., B\"acklund transformation for the {H}irota dif\/ference equation,
 and the supersymmetric {B}ethe ansatz, \href{https://doi.org/10.1007/s11232-008-0047-2}{\textit{Theoret. and Math. Phys.}}
 \textbf{155} (2008), 567--584.

\bibitem{Zakharov/Manakov}
Zakharov V.E., Manakov S.V., Construction of multidimensional nonlinear
 integrable systems and their solutions, \href{https://doi.org/10.1007/BF01078388}{\textit{Funct. Anal. Appl.}}
 \textbf{19} (1985), 89--101.

\bibitem{Zakharov/Shabat}
Zakharov V.E., Shabat A.B., A scheme for integrating the nonlinear equations of
 mathematical physics by the method of the inverse scattering problem.~I,
 \href{https://doi.org/10.1007/BF01075696}{\textit{Funct. Anal. Appl.}} \textbf{8} (1977), 226--235.

\end{thebibliography}
\end{document}